# Oxygen Deficient α-$MoO_3$ with Promoted Adsorption and State-Quenching of $H_2O$ for Gas Sensor: A DFT Study

Changmeng Huan[1,2], Pu Wang[1,2], Binghan He[1,2], Yongqing Cai[3]*, Qingqing Ke[1,2]*

[1]School of Microelectronics Science and Technology, Sun Yat-sen university, Zhuhai 519082, China

[2]Guangdong Provincial Key Laboratory of Optoelectronic Information Processing Chips and Systems, Sun Yat-sen University, Zhuhai 519082, China

[3]Joint Key Laboratory of the Ministry of Education, Institute of Applied Physics and Materials Engineering, University of Macau, Taipa, Macau, China

* Corresponding authors

E-mail: yongqingcai@um.edu.mo; keqingq@mail.sysu.edu.cn

## Abstract:

Semiconducting oxides with reducible cations are ideal platforms for various functional applications in nanoelectronics and catalysts. Here we report an ultrathin monolayer α-$MoO_3$ where tunable electronic properties and different gas adsorbing behaviors upon introducing the oxygen vacancies ($V_O$). The unique property of α-$MoO_3$ is that it contains three different types of oxygen atoms occupying three Wyckoff sites that are absent in other low-dimensional oxides and provides rich electronic hybridized states. The presence of $V_O$ triggers intermediate state in the gap at ~0.59 eV below the conduction band minimum and reduces the work function dramatically, together with new excitations at near infrared. The realigned Fermi level associated with the dangling state of $V_O$ reduces the neighboring Mo atoms and affects the gas adsorption thereafter. The binding energy of $H_2O$ molecules above $V_O$ is 2.5 times up to -0.75 eV compared with that of perfect lattice site and trends of transfer of electrons also reverse. The latter is related with the shallow localized state in the band gap due to $H_2O$ adsorbed above perfect $MoO_3$ which becomes quenched upon adsorbing at the $V_O$ site. Those rich in-gap defective states in oxygen deficient $MoO_3$, broadening the light absorption and promoting the uptake of water, are conductive to the application of α-$MoO_3$ for optoelectronics, photothermal therapy, and sensor of moisture.

**Keywords:** Monolayer α-$MoO_3$, Oxygen vacancy, $H_2O$ adsorption, Gas sensor, Optoelectronic

## 1. Introduction

Two-dimensional (2D) transition metal oxides (TMOs) have received considerable attention owing to their distinctive electronic properties as well as high specific surface area and rich active surface sites[1-6]. Orthorhombic phase α-$MoO_3$, a stable layered TMO, has been widely used as anode material for ion batteries (e.g., $H^+$, $Li^+$, $Zn^{2+}$) and supercapacitors due to its multiple valence states and unique layered structure[7-9]. Recently, anisotropic phonon polaritons and photonic magic angles were demonstrated in twisted α-$MoO_3$, offering unprecedented opportunities for controlling light at the nanoscale and making the ultrathin α-$MoO_3$ an interesting subject in materials research[10-13].

Significantly, electronic and optical properties of α-$MoO_3$ are highly tunable with oxygen vacancies ($V_O$). Associated with the enhanced electrical conductivity and additional active sites induced by $V_O$, $MoO_{3-x}$ electrodes exhibit superior rate performance and excellent cycling stability[14-16]. Furthermore, the work function of α-$MoO_3$ is tunable with the defective levels induced by $V_O$, which has been utilized to design the $MoS_2$/α-$MoO_{3-x}$ heterojunction as phototransistors with a high detectivity of $9.8 \times 10^{16}$ cm $Hz^{1/2}$ $W^{-1}$ [17]. Meanwhile, the defective levels can achieve near infrared (NIR) absorption and localized surface plasmon resonance (LSPR) effect, making α-$MoO_{3-x}$ promising as biodegradable nanoagents for photothermal cancer therapy[18, 19].

On the other hand, α-$MoO_3$ has been demonstrated to be a promising material for gas sensing because of its high sensitivity, fast response and thermodynamic stability[20]. Xu et al. have studied the adsorption behaviors of many gases (such as $H_2$, $H_2S$, $NH_3$, CO) molecules on the surface of $MoO_3$, providing a conceptual foundation for $MoO_3$-based gas sensor[21, 22]. In addition, adsorbates from ambient conditions may cause degradation and affect the remarkable properties of the two-dimensional material, especially in the presence of vacancies[23-26]. Unfortunately, the effects of the $V_O$ and its associated coupling with $H_2O$ molecules in α-$MoO_3$ have been scarcely reported, and understanding the interplay of the structural defect and adsorbates is prerequisite for its massive applications.

In this work, we systematically studied the adsorption behavior of $H_2O$ molecules on the surface and oxygen vacancies of monolayer α-$MoO_3$ through density functional theory (DFT) calculations, as well as the electronic properties, interfacial charge transfer, work function, and optical properties. Our results indicates that the adsorption of $H_2O$ molecules on the surface of α-$MoO_3$ will introduce an intermediate state close to the edge of valence band and allowing the thermally activated conduction, which is necessary for monitoring relative humidity in moisture-sensitive environment. For oxygen deficient α-$MoO_3$, the depletion of O atoms makes the dangling Mo atoms reduced and promotes the adsorption of $H_2O$. While the $H_2O$ donates electrons to $MoO_3$ for adsorbing above perfect lattice, it accepts electrons when adsorbing above $V_O$ site. This strong interaction between $V_O$-$H_2O$ complex makes oxygen deficient $MoO_{3-x}$ a promising moisture sensor. Moreover, the rich in-gap defective states in oxygen deficient $MoO_3$ can modulate the work function and carrier density, broaden the light absorption, and promote the uptake of water, which highlights the prospect of α-$MoO_3$ as tunable optoelectronics, photothermal and sensing materials.

## 2. Computational method and details

The first-principles calculations based on the projected augmented wave (PAW) method as implemented in Vienna Ab Initio Simulation Package (VASP) were employed[27, 28]. The Perdew-Burke-Ernzerhof (PBE) functional of the generalized gradient approximation (GGA) was used to treat the exchange and correlation[29]. Since the pure DFT is less accurate to describe van der Waals (vdW) interactions, the geometric structures were fully relaxed through PBE functional with vdW-D3 correction[30, 31]. The plane-wave cutoff energy of 400 eV and a Monkhorst-Pack k-points sampling of 3 × 3 × 1 for the 4 × 4 × 1 monolayer supercell were adopted. A vacuum layer of 15 Å in z direction was inserted to avoid mirror interactions. The structural optimizations were considered to be converged when the system energies and Hellman-Feynman force on each atom were less than $1\times10^{-6}$ eV and 0.01 eV/Å, respectively. Since the PBE method seriously underestimates the band gap, we utilized the standard Heyd-Scuseria-Ernzerhof (HSE06) hybrid functional to calculate the electronic structure and optical properties[32]. The effective band structures were calculated by the corresponding module in VASPKIT[33].

In Van de Walle method, the formation energy ($E_f$) of oxygen vacancies is calculated through[34]

$$E_f = E_d - E_p + n_O \mu_O \qquad (1)$$

where $E_d$ and $E_p$ represent the total energies of the defective and perfect α-$MoO_3$, respectively, $n_O$ indicates the number of $V_O$, and $\mu_O$ is the chemical potential of O atom. The range of $\mu_O$ in $MoO_3$ can be expressed as[35]

$$\mu_O(\text{Mo-rich}) < \mu_O^{MoO_3} < \mu_O(\text{O-rich}) \qquad (2)$$

where $\mu_O$(O-rich) represent the oxygen chemical potentials equals to $\mu_O^O$ of $O_2$ molecular (per oxygen atom). $\mu_O$(Mo-rich )=$\mu_O^O + \Delta E_f^{MoO_3}$ corresponds to the Mo-rich conditions and the $\Delta E_f^{MoO_3}$ is the formation energy of $MoO_3$ from metal Mo and $O_2$ gas. If the charged defects are considered, the formation energy ($E_f$) formula should be extended as[36]

$$E_f = E_d - E_p + n_O \mu_O + q\,( E_F + E_V + \Delta V ) \qquad (1)$$

where $E_F$ is the Fermi level referenced to the energy of the VBM ($E_V$) in perfect $MoO_3$, ranging from 0 to the value of the band gap. A correction term $\Delta V$ was added to align the reference electrostatic potential in defective $MoO_3$ with that in perfect $MoO_3$.

The adsorption energies ($E_{ads}$) of $H_2O$ molecule on α-$MoO_3$ systems are calculated as[37]

$$E_{ads} = E_{i+w} - E_i - E_w \qquad (3)$$

where $E_{i+w}$, $E_i$, and $E_w$ represent the energies of $H_2O$ adsorbed α-$MoO_3$, the pristine α-$MoO_3$, and the isolated $H_2O$ molecule, respectively.

## 3. Results and Discussion

**Pristine perfect α-$MoO_3$ monolayer:**

The crystal structure of bulk α-$MoO_3$ (Fig. 1a) belongs to the orthorhombic *Pbnm* space group, which consists of alternating layers with corner-sharing and distorted $MoO_6$ octahedrons. There are merely weak vdW interactions between adjacent monolayers, therefore the monolayer α-$MoO_3$ (Fig. 1b) can be acquired through mechanical cleavage along the (010) plane. As marked in Fig. 1c and d, the distorted $MoO_6$ octahedron has three distinct oxygen sites: terminal oxygen ($O_t$), asymmetric oxygen ($O_a$), and symmetric oxygen ($O_s$). In the optimized bulk α-$MoO_3$, the $O_t$ atoms is exclusively coordinated with one Mo atom with a bond length of 1.70 Å. The $O_a$ atoms are 2-fold coordinated forming a relatively long (2.21 Å) and short (1.77 Å) bonds with Mo atoms in the same layer. The $O_s$ atoms are tri-coordinated forming two equal bonds (1.95 Å) with Mo atoms in the same layer and a much longer bond (2.38 Å) with one Mo atom in the other sublayer. As shown in Table 1, the calculated lattice parameters and M-O bond length for bulk $MoO_3$ in this work are in line with those of previous theoretical work and experiment[21, 38]. In the monolayer case, the bond (2.47 Å) between $O_s$ and Mo atoms in the other sublayer is slightly longer than that in bulk of 2.38 Å. This reflects the relaxation of the lattice in monolayer due to the disappearing of the interlayer interactions as presence in bulk α-$MoO_3$.

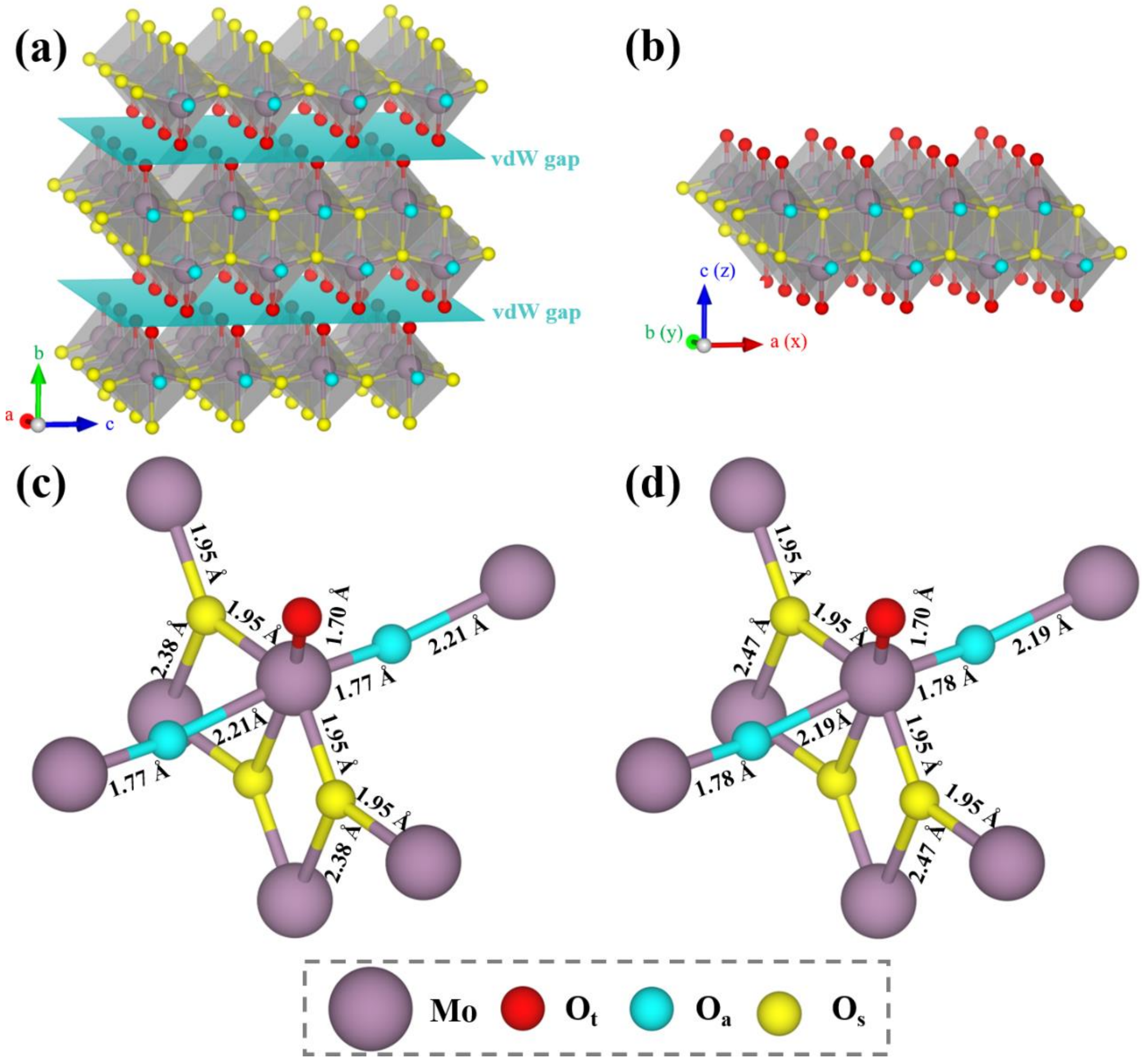

**Fig. 1** Structure of the optimized bulk α-$MoO_3$ supercell (a) along the [010] orientation and monolayer α-$MoO_3$ supercell (b). The distorted $MoO_6$ octahedrons with nearest neighbor Mo atoms in bulk (c) and monolayer (d) highlighting the oxygen sites and bond length.

**Table 1.** Calculated lattice parameters and M-O bond length for bulk and monolayer α-

$MoO_3$ compared to previous theoretical work and experiment.

| Structure | Bulk | Monolayer Supercell | Bulk[21] (Previous work) | Bulk[38] (Experiment) |
|---|---|---|---|---|
| a (Å) | 3.96 | / | 3.96 | 3.96 |
| b (Å) | 13.86 | / | 13.86 | 13.86 |
| c (Å) | 3.70 | / | 3.70 | 3.70 |
| $d_{Mo\text{-}Ot}$ (Å) | 1.70 | 1.70 | 1.68 | 1.67 |
| $d_{Mo\text{-}Oa\ 1}$ (Å) | 1.77 | 1.78 | 1.75 | 1.73 |
| $d_{Mo\text{-}Oa\ 2}$ (Å) | 2.21 | 2.19 | 2.24 | 2.25 |
| $d_{Mo\text{-}Os\ 1}$ (Å) | 1.95 (2×) | 1.95 (2×) | 1.95 (2×) | 1.95 (2×) |
| $d_{Mo\text{-}Os\ 2}$ (Å) | 2.38 | 2.47 | 2.38 | 2.33 |

**Energetics of oxygen vacancies in monolayer α-$MoO_3$:**

Transition metal oxides have a tendency to form oxygen vacancies which have profound influence on their physical and chemical properties[39, 40]. In the monolayer α-$MoO_3$ supercell, three types of oxygen vacancies can therefore be considered and calculated through removing the oxygen atoms lying on the terminal, asymmetric and symmetric sites. The formation energies of three types of $V_O$ and geometry structure of oxygen deficient $MoO_3$ systems were calculated and presented in Table 2 and Fig. 2, respectively. We found that both $V_{Ot}$-$MoO_3$ and $V_{Oa}$-$MoO_3$, showing similar structure after relaxation and same energy, are around 2.15 eV lowered in total energy than that of the $V_{Os}$-$MoO_3$. This indicates that $V_{Ot}$ and $V_{Oa}$ are more energetically favored compared with $V_{Os}$-$MoO_3$. As shown in Fig. 2(a-c), the formation of $V_O$ leads to a strong geometrical rearrangement around $V_O$. For $V_{Ot}$-$MoO_3$, the nearest $O_a$ atom moves towards the $V_{Ot}$ while the dangling Mo atom downshift, causing the bond length of Mo-$O_a$ to decrease from 2.19 and 1.78 Å to 1.80 and 1.70 Å, respectively. Meanwhile, bond length of the dangling Mo with the $O_s$ in the other sublayer reduces from 2.47 to 2.18 Å. More importantly, our results demonstrate that $V_{Ot}$- and $V_{Oa}$-$MoO_3$ show similar geometry structure (the remaining O atom stays at an intermediate position between initial $O_t$ and $O_a$ sites) and the same formation energies after relaxation. The negative $E_f$ of $V_O$ in Mo-rich (O-poor) conditions indicates that the formation of $V_O$ in a reducing atmosphere or ultra-high vacuum is energetically favored, which is consistent with the similar trends of $V_O$ in wide bandgap semiconductors[41]. Our work suggests that the production of the $V_O$ spans a broad range of the oxygen chemical potential which allows flexible modulation and control of its content and distribution such defects via external pressures and temperatures.

**Table 2.** The relative energies ($\Delta E_0$) of different $MoO_3$ systems and the formation energies of different oxygen vacancies in PBE method, where the total energy of $V_{Os}$-$MoO_3$ is set to zero for comparison

| | $V_{Ot}$-$MoO_3$ | $V_{Oa}$-$MoO_3$ | $V_{Os}$-$MoO_3$ |
|---|---|---|---|
| $\Delta E_0$ (eV) | -2.15 | -2.15 | 0 |
| $E_f$(Mo-rich) (eV) | -5.72 | -5.72 | -3.57 |
| $E_f$(O-rich) (eV) | 2.08 | 2.08 | 4.23 |

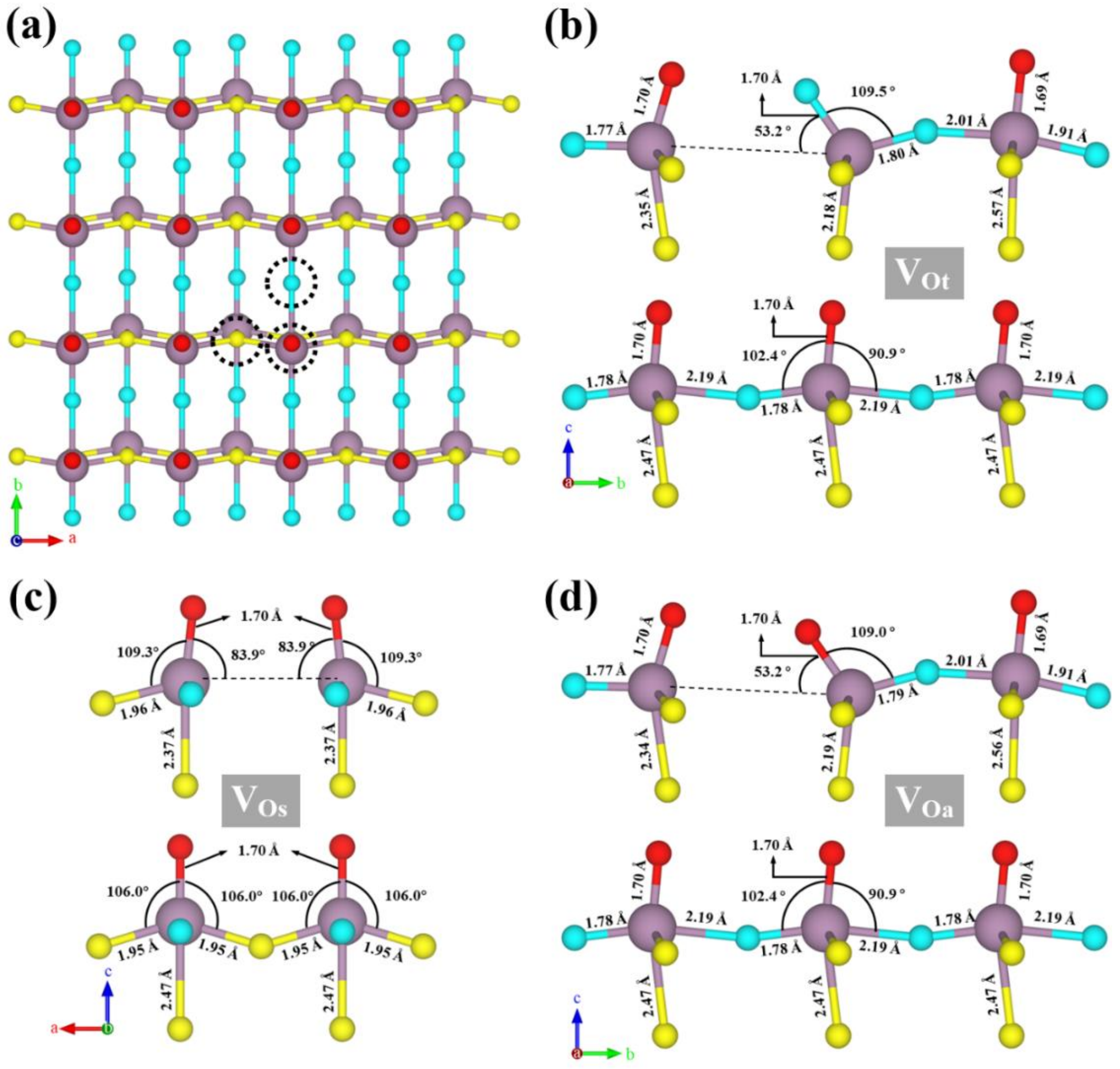

**Fig. 2** (a) Top view of the monolayer α-$MoO_3$ supercell marked with three types of oxygen sites. The structural features of the optimized and pristine $V_{Ot}$-$MoO_3$ (b), $V_{Os}$-$MoO_3$ (c), and $V_{Oa}$-$MoO_3$ (d) system.

For analysis and comparison, charged oxygen vacancies are also considered. The results in Fig. S1 and Table. S1 show that $V_O^0$ are more thermodynamically favored than $V_O^{1+}$ and $V_O^{2+}$ in oxygen deficient α-$MoO_3$ (n-type semiconductor). Furthermore, the adsorption behavior, electronic and optical properties of $V_{Ot}$-$MoO_3$ and $V_{Oa}$-$MoO_3$ (Fig. S2) are the same. Therefore, only charge-free $V_{Ot}$-$MoO_3$ will be considered in the following discussions.

**Adsorption of $H_2O$ molecule:**

The adsorbates from ambient conditions may influence the electronic and optical properties of 2D materials, herein, the adsorption behavior of $H_2O$ on the surface of α-$MoO_3$ systems is considered. $H_2O$ molecules are polar molecules, with negative charge (σ-) on the O atom side and positive charge (σ+) on H atom side. According to the principle of mutual attraction between (σ+) and (σ-), five possible adsorption sites of $H_2O$ above perfect $MoO_3$ are considered in Fig. S3. In site 1, one arm of H-O-H vertically pointing to a $O_t$ of $MoO_3$. In site 2 and 3, the H-O-H are located above the center of the two $O_t$ atoms along the $O_a$ and $O_s$ arrangement direction, respectively. In site 4 and 5, the H-O-H atoms are located diagonally above the center of the rectangle formed by four $O_t$ atoms with O-H bonds pointing up and down, respectively. Based on the lowest-energy configuration, the total energies and adsorption energies of final structure are summarized in Table S2. The most stable configuration ($E_{ads}$ of -0.30 eV) is at site 4, where the oxygen of the water slightly deviates above the center of the rectangle formed by four $O_t$ atoms with two O-H bonds pointing up. The uptake of more water molecules on the surface were also calculated. To compare the trend, we examined higher coverages with the adsorption of 1 to 4 $H_2O$ molecules on the surface of α-$MoO_3$, and the results are shown in Table 3. The adsorption energies increase almost linearly with the increase in the number of $H_2O$, and the average adsorption energy per molecule $H_2O$ remains basically unchanged. As shown in Fig. 3 (a), the structure and parameters of 4-$H_2O$ adsorbed α-$MoO_3$ systems are basically consistent with 1-$H_2O$ adsorbed α-$MoO_3$ (Fig. S3 site 4), which means a strong adsorption capacity on the surface of monolayer α-$MoO_3$. These coplanar and equally separated of adsorption of adsorbed $H_2O$ would favor the proton hopping in hydrated $MoO_3$. The shallow defective states arisen from $H_2O$ in the band gap of $MoO_3$ also alter the behaviors of charge dynamics as we shown below.

**Table 3.** The adsorption energies ($E_{ads}$) of perfect and defective α-$MoO_3$ systems, where $E_{ads\text{-}A}$ represents the average adsorption energies per molecule $H_2O$.

| | 1-$H_2O$ | 2-$H_2O$ | 3-$H_2O$ | 4-$H_2O$ | $V_O$-$H_2O$ |
|---|---|---|---|---|---|
| $E_{ads}$ (eV) | -0.30 | -0.60 | -0.87 | -1.15 | -0.75 |
| $E_{ads\text{-}A}$ (eV) | -0.30 | -0.30 | -0.29 | -0.29 | -0.75 |

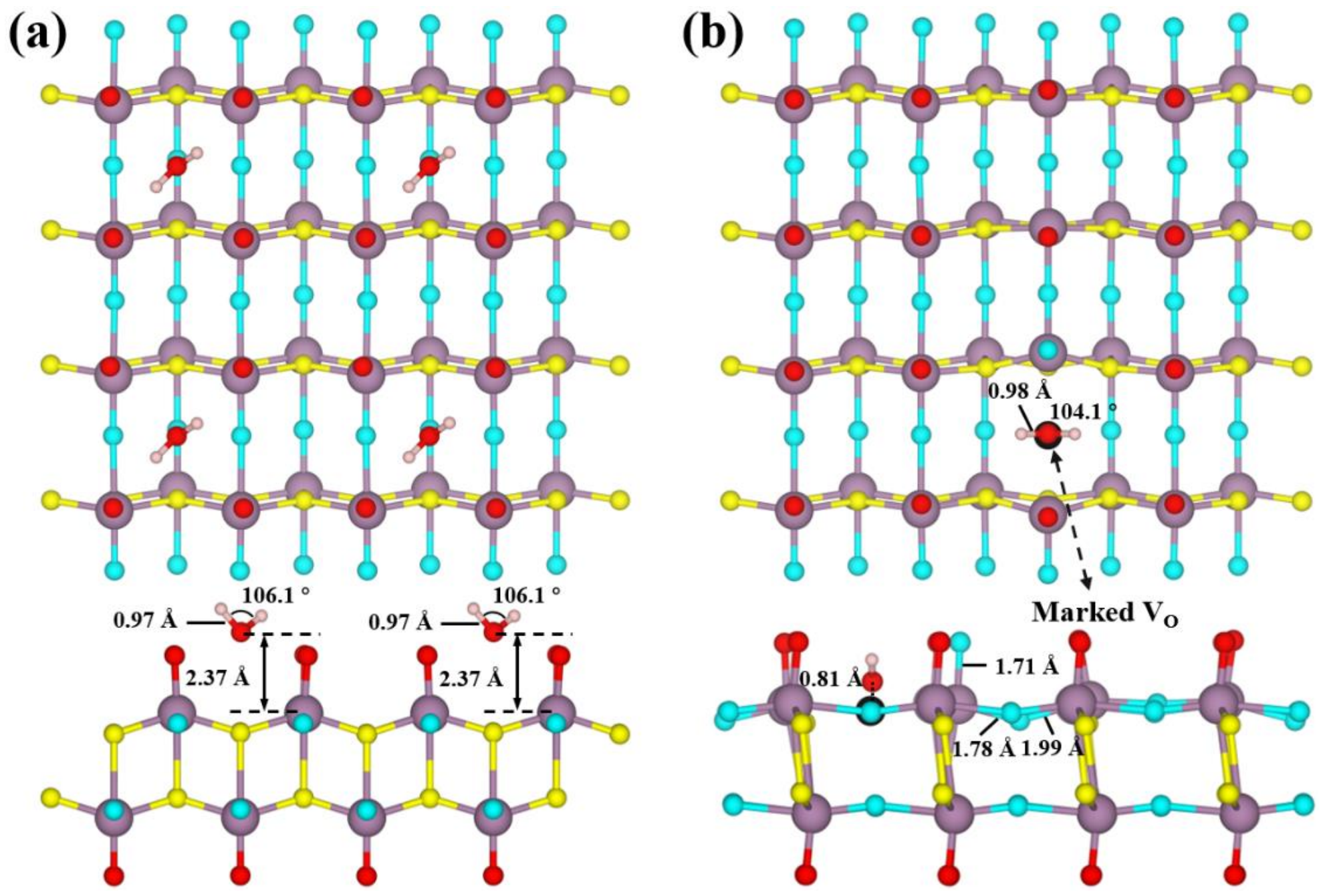

**Fig. 3** Top and side views of the optimized structures of 4-$H_2O$ molecules on the surface of perfect-$MoO_3$ (a) and $H_2O$ molecule on the surface of $V_{Ot}$-$MoO_3$ (b). The red and pink balls represent O and H atoms, respectively.

To study the adsorption behavior of $H_2O$ molecule at $V_O$, we obtained the optimized structure from an initial structure in which $H_2O$ molecule is located at the marked $V_O$ site (Fig. 3b). As shown in Table 3, the presence of $V_O$ on the surface significantly promotes the adsorption of $H_2O$ molecules. The $E_{ads}$ for $V_{Ot}$-$MoO_3$ (-0.75 eV) is 2.5 times the values for perfect-$MoO_3$ with lowest configurations at site 4 (-0.30 eV). In addition, in Fig. 3 (b), the adsorption of $H_2O$ also restores the bond lengths of certain Mo-O bonds in $V_O$-$MoO_3$ toward the values for perfect-$MoO_3$. This suggests that the adsorbed $H_2O$ could partially passivate the $V_O$.

The relatively large dipole moment of $H_2O$ is liable to induce a charge redistribution in the adsorbed α-$MoO_3$ surface. To analyze the electronic interaction between the $H_2O$ molecule with the $MoO_3$, we conducted a calculation on the differential charge density (DCD) $\Delta\rho$(r) defined as the difference between the total charge density of $H_2O$ adsorbed $MoO_3$ system minus the sum of the charge density of the isolated $H_2O$ molecule and $MoO_3$ system (perfect-$MoO_3$ or $V_O$-$MoO_3$). To obtain the exact amount of transferred charge, the plane-averaged DCD $\Delta\rho$(z) along the normal direction (*z*) direction of $MoO_3$ system is calculated by integrating $\Delta\rho$(r) within the basal plane at the z point. The amount of transferred charge at z point is given by $\Delta Q(z) = \int_{-\infty}^{z} \Delta\rho(z') \, dz'$ [42, 43]. Based on the $\Delta Q(z)$ curves, the total amount of charge donated by the molecule is read at the interface between the $H_2O$ and the $MoO_3$ system, where $\Delta\rho$(z) shows a zero value.

The isosurfaces of $\Delta\rho(r)$ for the $H_2O$ molecule adsorbed on perfect $MoO_3$ and $V_O$-$MoO_3$ are plotted in Fig. 4, respectively. Fig. 4a shows that there is a depletion of electrons in $H_2O$ molecule and an accumulation of electrons in the nearest O atoms of perfect $MoO_3$ surface, and the $H_2O$ molecule donates electrons to α-$MoO_3$ with 0.02 e per molecule. For the $MoO_3$ surface decorated with four $H_2O$ molecules, the electrons transfer from $H_2O$ molecules to α-$MoO_3$ is up to 0.09 e with about 0.02 e per molecule, which is consistent with the linear increase of adsorption energy in Table 3. This trend is same to the slightly positively charge of $H_2O$ above phosphorene but different from those adsorptions above graphene and antimonene[44]. In contrast, for $H_2O$ adsorbing above $V_O$ ($V_O$-$MoO_3$-$H_2O$), there is an accumulation of electrons in $H_2O$ molecule and a depletion of electrons in the nearest Mo atoms of $V_O$-$MoO_3$, and the $H_2O$ molecule accepts electrons from $V_O$-$MoO_3$ with around 0.06 e per molecule (Fig. 4b), which results from the occupied Mo $4d_{xy}$ gap state introduced by $V_O$ (see next section). The charge transfer from this localized state induced by $V_O$ to the $H_2O$ molecule is more likely to occur[45]. Therefore, the presence of $V_O$ could alter the energetics and charged state of adsorbed $H_2O$. The presence of $H_2O$ and $V_O$ may significantly affect the properties such as carrier density and mobility which is appealing for sensor of moisture.

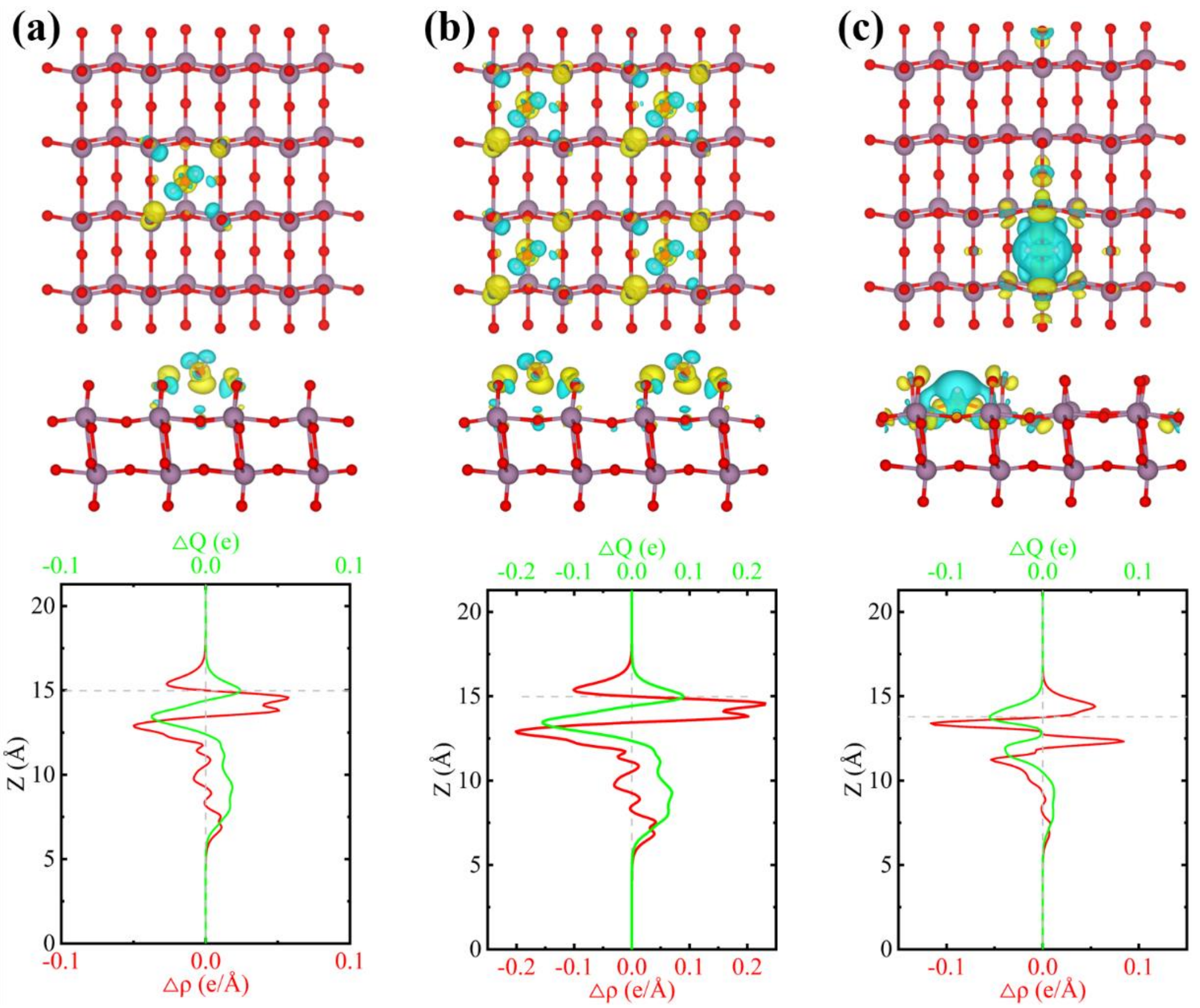

**Fig. 4** Charge redistribution for $MoO_3$-$H_2O$ (a), $MoO_3$-$4H_2O$ (b), and $V_O$-$MoO_3$-$H_2O$ (b). Top and middle panels: Top and front views of the DCD isosurfaces, where the cyan (yellow) region represents depletion (accumulation) of electrons. Bottom panel: Plane-averaged DCD $\Delta\rho(z)$ (red line) and the amount of transferred charge $\Delta Q$(z) (green line) between the $H_2O$ molecule and $MoO_3$.

**$V_O$ and $H_2O$ induced changes of electronic states:**

The band structures of perfect-$MoO_3$, perfect-$MoO_3$ adsorbed $H_2O$ ($MoO_3$-$H_2O$), $V_O$-$MoO_3$, and $V_O$-$MoO_3$-$H_2O$ were calculated using HSE06 method. The CBM and valence band maximum (VBM) of Perfect-$MoO_3$ are 9.60 eV and 6.67 eV below the vacuum (Fig. 5a), respectively, together with an indirect bandgap of 2.92 eV, which are consistent with the previous results obtained by ultraviolet photoemission spectroscopy (UPS) and inverse photoemission spectroscopy (IPES) in bulk[46-48]. Interestingly, the adsorption of $H_2O$ molecule introduces an intermediate gap state at ~0.30 eV above the VBM (Fig. 5b) and maintain a band gap of 2.92 eV consistent with perfect-$MoO_3$. The relatively shallow state would trap the holes and alter the carrier dynamics in $MoO_3$. The $H_2O$ molecule induced intermediate gap state in $MoO_3$ has not been found in other 2D systems like phosphorene and antimonene[44], where there is no water induced states in the band gap.

For oxygen deficient $MoO_3$, the formation of $V_{Ot}$ introduces a gap state at ~0.59 eV below the CBM and reduces the bandgap from 2.92 to 2.81 eV, which is consistent with experimental result[49]. This is due to the fact that the electrons transfer from the $V_O$ to the 4d states of Mo sufficiently reduces the crystal field[50]. Absorbingly, the intermediate state associated with $H_2O$ above perfect $MoO_3$ disappears for $H_2O$ adsorbed above Vo-$MoO_3$, leaving an intermediate state at ~0.57 eV below the CBM in Fig. 5d. These intermediate gap states in $V_O$-$MoO_3$ and $V_O$-$MoO_3$-$H_2O$ could be responsible for the broadened window of light-absorption[51].

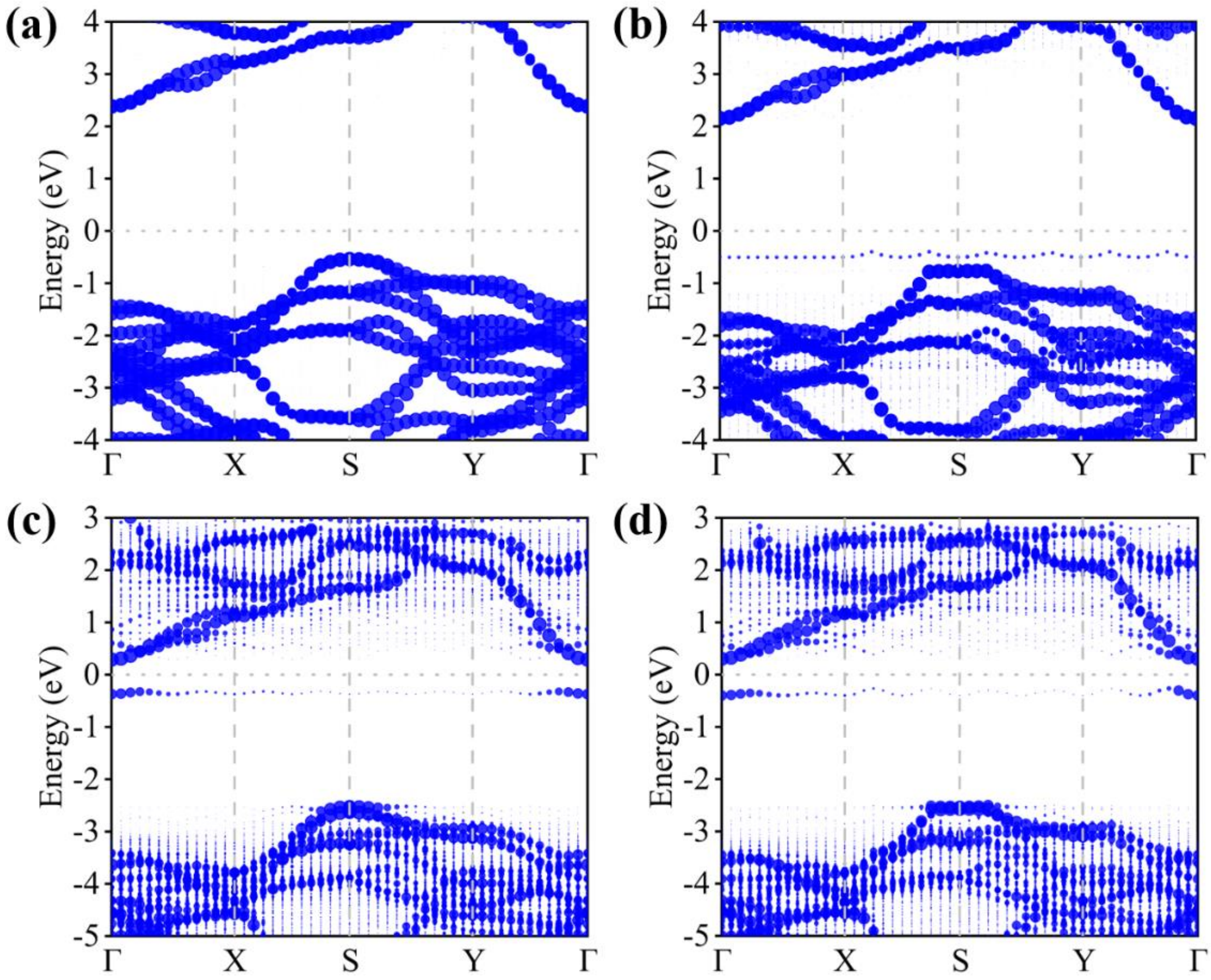

**Fig. 5** Effective band structures for different α-$MoO_3$ systems acquired from HSE06 method, (a) perfect-$MoO_3$, (b) $MoO_3$-$H_2O$, (c) $V_O$-$MoO_3$, and (d) $V_O$-$MoO_3$-$H_2O$. The Fermi energy is set to zero and marked with a horizontal dashed line.

The projected density of states (PDOS) of different α-$MoO_3$ systems obtained from HSE06 method are shown in Fig. 6. In all systems, the top of the valence band (VB) is mainly composed of O 2p orbitals, while the bottom of the conduction band (CB) is characterized by Mo 4d states, which are in good agreement with previous study[50]. Comparing Fig. 6a and b, it can be found that an intermediate state appears near the VBM after adsorbing $H_2O$ which is mainly attributed to the $2p_x$ and $2p_y$ orbitals of O in $H_2O$, which increases the contribution to the VBM. Furthermore, as the number of adsorbed $H_2O$ molecules increases, the water-induced shallow state becomes stronger and extends to the edge of valence band (Fig. 6c), facilitating a promoted conductivity of the systems, which is necessary for $H_2O$ sensing. The same phenomenon occurs in $NO_2$ adsorbed silicon oxycarbide[52].

In Fig. 6d and e, there are occupied gap states near the CBM in both the $V_O$-$MoO_3$ and $V_O$-$MoO_3$-$H_2O$, which serve as donor levels in the bandgap, giving rise to a significant up-shift in the Fermi level and making the systems n-type. The formation of $V_{Ot}$ donates 2e to the Mo-4d states, which reduces the crystal field and results in the formation of an extended $4d_{xy}$ gap state (inset in Fig. 6) [20]. Consistent with the effective band structure (Fig. 5), the $H_2O$ related shallow state quenches and disappears upon $H_2O$ adsorbing above $V_O$. Interestingly, compared with the naked $V_O$-$MoO_3$, the addition of $H_2O$ effectively promotes the DOS at the band edges of CB and VB. This in turn will increases the number of charged carriers under excitation.

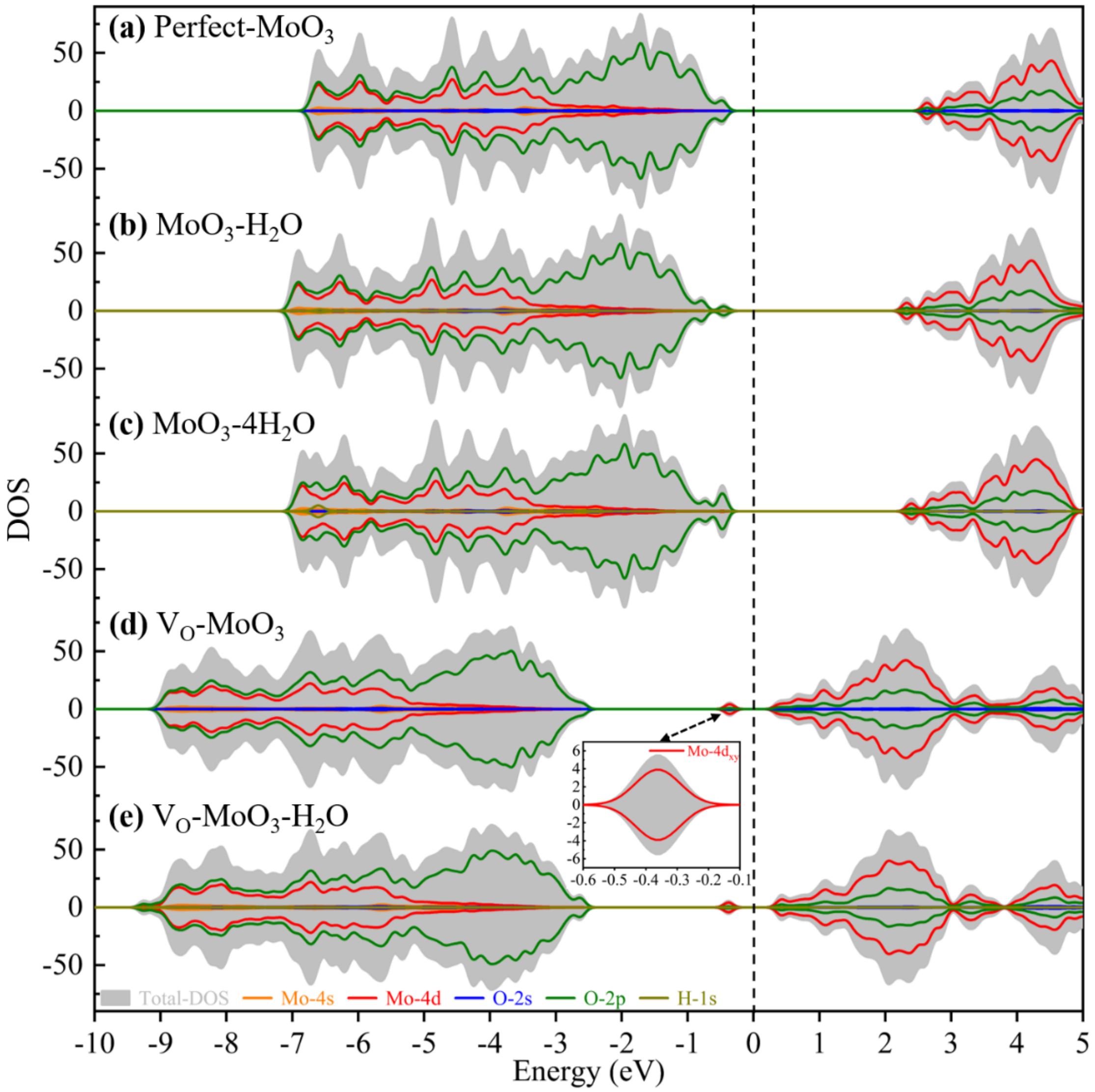

**Fig. 6** PDOS of perfect-$MoO_3$ (a), $MoO_3$-$H_2O$ (b), $MoO_3$-4$H_2O$ (c), $V_O$-$MoO_3$ (d), and $V_O$-$MoO_3$-$H_2O$ (e). Inset shows the enlargement of the defective state associated with $V_O$. The Fermi level is set as 0 and marked with a vertical dashed line.

**Optical response of oxygen deficient surface and effect of moisture:**

The optical absorption is crucial for the optoelectronic and photothermal application, while defect-free α-$MoO_3$ suffers from limited range of light absorption. Therefore, we conducted calculations on the optical absorption spectra of perfect-$MoO_3$, $MoO_3$-$H_2O$, $V_O$-$MoO_3$, and $V_O$-$MoO_3$-$H_2O$, and plotted the results in Fig. 7a and b. All $MoO_3$ systems exhibit anisotropic optical absorption. As shown in Fig. 7a, both the perfect $MoO_3$ and $MoO_3$-4$H_2O$ merely absorb ultraviolet (UV) light. However, there is a slight red-shift (inset of Fig. 7a) of absorption spectra with the increase of adsorbed $H_2O$ molecules, which is due to the intermediate gap state extending to the edge of valence band (Fig. 6b and c). Meaningfully, there are strong NIR absorption from 0.5 to 1.3 eV and weak visible light absorption in $V_O$-$MoO_3$ and $V_O$-$MoO_3$-$H_2O$ (Fig. 7b), which are attributed to the excitations from intermediate state to CBM and VBM to intermediate state, respectively. Our calculation results may explain the previous experimental results where promoted light adsorption and LSPR effect by introduction of $V_O$ for photo-thermal synergistic catalytic and photothermal therapy[53-55]. As shown in the schematic diagrams in Fig. 7c and d, the perfect-$MoO_3$ and $MoO_3$-$H_2O$ only absorb UV light through excitations from the VBM to CBM. In contrast, α-$MoO_3$ with $V_O$ can not only absorb UV light but also absorb NIR light through the intermediate state with excitation from Mo-4$d_{xy}$ to O-2p empty levels. Hence, the optical properties of α-$MoO_3$ can be modulated by proper introduction of $V_O$ to achieve applications in photocatalysis, photothermal therapy, and photodetector.

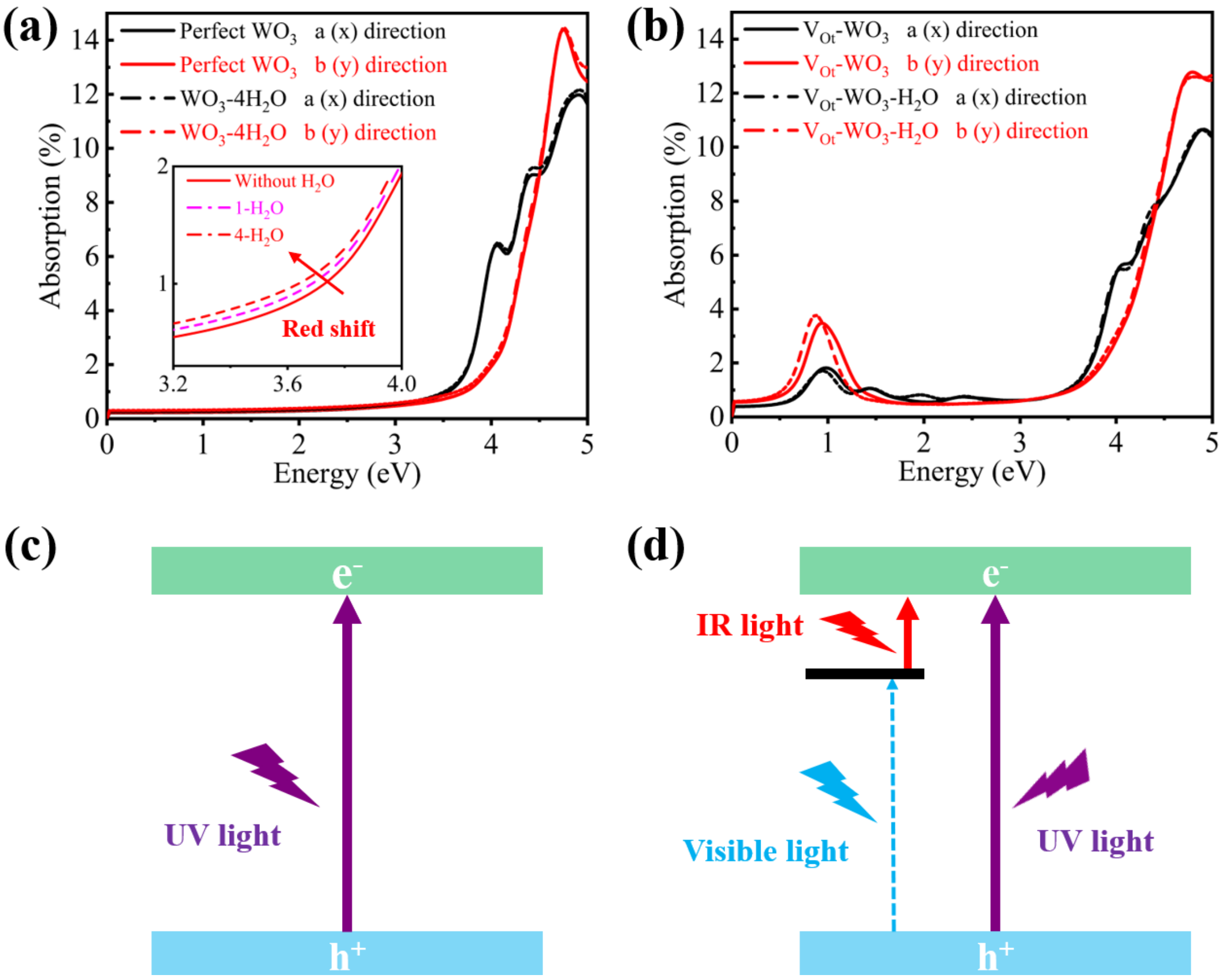

**Fig. 7** The absorption spectra of perfect-$MoO_3$ and $MoO_3$-$4H_2O$ (a), $V_O$-$MoO_3$ and $V_O$-$MoO_3$-$H_2O$ (b). Schematic energy band diagram of optical absorption in $MoO_3$-$H_2O$ (c) and $V_O$-$MoO_3$-$H_2O$ (d). The inset in (a) show a red-shift of absorption spectra with the increase of $H_2O$ molecules.

**Variation of work function with oxygen deficiency and $H_2O$ adsorption:**

Fig. 8 summarizes the alignment of the VBM and CBM with respect to the vacuum level for different systems based on the HSE06 calculation. The difference between CBM and VBM of different systems is slight, but the position of the Fermi level is quite different. The positions of the Fermi level are -6.96, -5.64, -3.93 and -3.85 eV for perfect-$MoO_3$, $MoO_3$-$H_2O$, $V_O$-$MoO_3$, and $V_O$-$MoO_3$-$H_2O$, respectively. The work functions were also calculated accordingly based on $\varphi = E_{vac} - E_F$, where $\varphi$, $E_{vac}$, and $E_F$ represent the work function, vacuum level, and Fermi level, respectively. The calculated $\varphi$ are 9.07, 8.74, 7.03, and 6.96 eV for perfect-$MoO_3$, $MoO_3$-$H_2O$, $V_O$-$MoO_3$, and $V_O$-$MoO_3$-$H_2O$, respectively. The calculated work function of $V_O$-$MoO_3$, and $V_O$-$MoO_3$-$H_2O$ are in good agreement with the experimental result obtained by UPS, where the work function and Fermi level of near-stoichiometric bulk $MoO_{3-x}$ are 6.89 eV and 0.39-0.59 eV below the CBM, respectively[48]. Furthermore, the introduction of intermediate state can effectively enhance the conductivity of $V_O$-$MoO_3$ or $MoO_{3-x}$ which reduces the Fermi energy to slightly above the VBM of the active materials in solar cells, and make oxygen deficient $MoO_{3-x}$ promising as a hole transport layer in solar cells and LEDs[56, 57]. Accordingly, introduction of $V_O$ dramatically modulates the work function, which in turn can affect the height of the Schottky barrier or formation of Ohmic contact in practical applications. In ultrathin monolayer or few-layer $MoO_3$, the weak electronic screening may imply an appreciable width of the depletion/accumulation regions. In contrast, the adsorption of $H_2O$ makes a weak effect on the work function, which also forms an intermediate state and marginally shifts the position of the CBM and VBM.

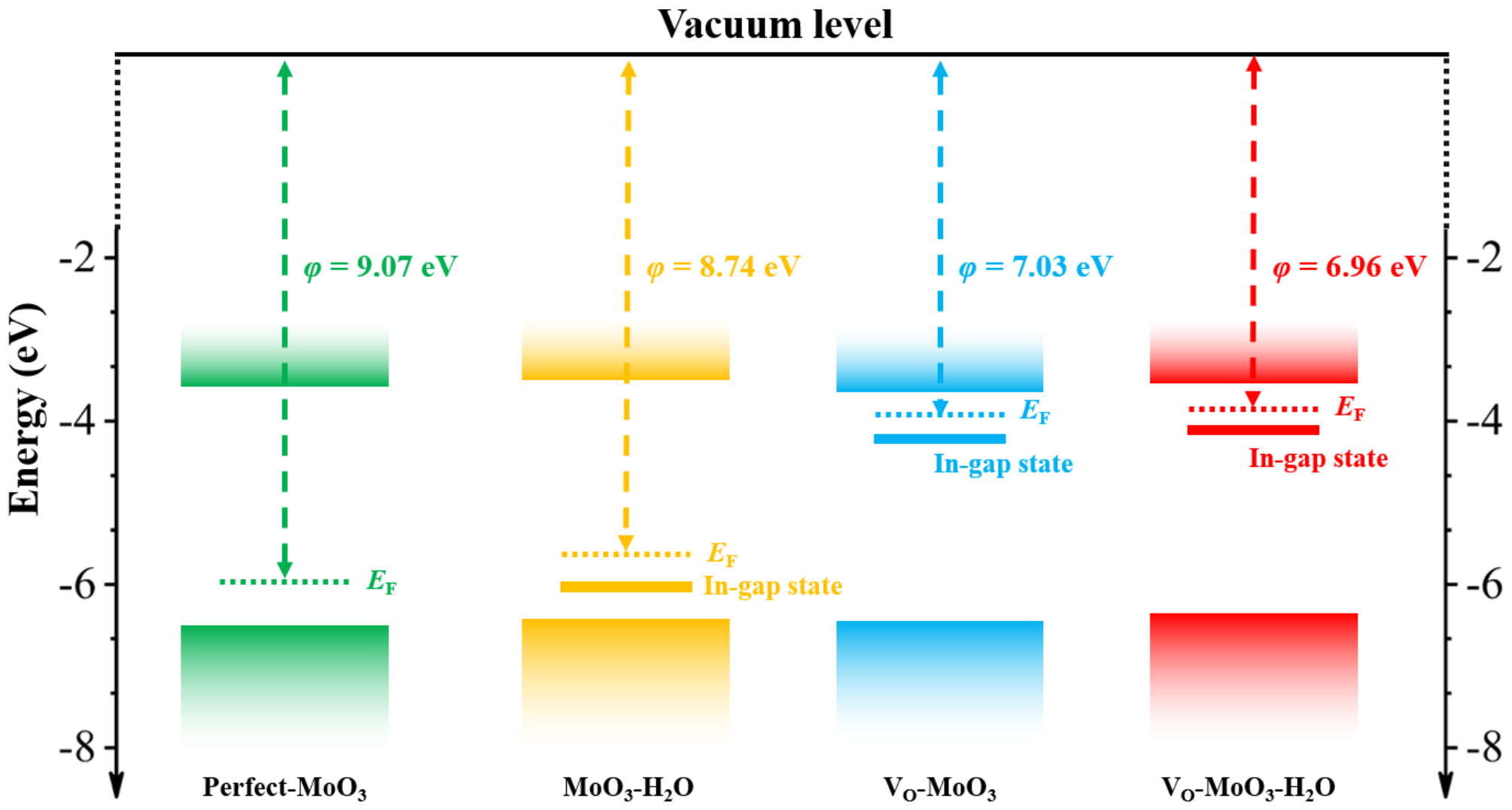

**Fig. 8** Work functions of perfect-$MoO_3$, $MoO_3$-$H_2O$, $V_O$-$MoO_3$, and $V_O$-$MoO_3$-$H_2O$ calculated by HSE06 method. The Fermi level is marked with a horizontal dashed line.

## 4. Conclusion

In conclusion, we have examined the adsorption behavior of $H_2O$ molecules on the surface and oxygen vacancies of monolayer α-$MoO_3$, as well as the electronic properties, interfacial charge transfer, work function, and optical properties. Our results indicates that the adsorption of $H_2O$ molecules on the surface of α-$MoO_3$ will introduce an intermediate state close to the edge of valence band and allowing the thermally activated conduction, which is necessary for monitoring relative humidity in moisture-sensitive environment. For oxygen deficient α-$MoO_3$, oxygen vacancies at the terminal and asymmetric sites are more energetically favored and those vacancies create defective localized levels slightly beneath the CBM. The depletion of oxygen atoms makes the dangling Mo atoms reduced and promotes the adsorption of $H_2O$. Interestingly, the $H_2O$ induced shallow level slightly above VBM of perfect $MoO_3$ is quenched when $H_2O$ is anchored at the $V_O$ site. While the $H_2O$ donates electrons to $MoO_3$ for adsorbing above perfect lattice, it receives electrons when adsorbing above $V_O$ site. This strong interaction between $V_O$-$H_2O$ complex suggests that oxygen deficient $MoO_3$ could be a promising moisture sensor which can modulate the performance of conductivity and optoelectronics. Our work reveals that the presence of $V_O$ and $H_2O$ could alter the work function, the carrier density and optical response which are conducive to the application in optoelectronics, sensing, and photothermal therapy.

## Conflicts of interest

The authors declare that they have no known competing financial interests or personal relationships that could have appeared to influence the work reported in this paper.

## Acknowledgments

The authors acknowledge the funding support from the 100 Talents Program of Sun Yat-sen University (Grant 76220-18841201), the University of Macau (SRG2019-00179-IAPME) and the Science and Technology Development Fund from Macau SAR (FDCT-0163/2019/A3), the Natural Science Foundation of China (Grant 22022309) and Natural Science Foundation of Guangdong Province, China (2021A1515010024).